\newif\ifpretty
\prettytrue

\ifpretty
\documentclass[prl,twocolumn,preprintnumbers,amsmath,amssymb,a4paper,twoside]{revtex4}
\else
\documentclass[prl,preprint,preprintnumbers,amsmath,amssymb]{revtex4}
\fi

\newif\ifpdf
        \ifx\pdfoutput\undefined
        \pdffalse 
        \else
        \pdfoutput=1 
        \pdftrue
        \fi
\ifpdf
   \usepackage[pdftex]{graphicx}
   \DeclareGraphicsExtensions{.pdf, .png, .jpg}
   \graphicspath{{./FIG/}}
   \usepackage{thumbpdf}
\else
   \usepackage{graphicx}
   \DeclareGraphicsExtensions{.eps}
   \graphicspath{{./FIG/}}
\fi

\usepackage{longtable}
\begin{document}

\title{Characterizing the stepwise transformation from a low-density to
  a very-high-density form of supercooled liquid water}
\author{Dietmar Paschek}
\email{dietmar.paschek@udo.edu}
\author{Alfons Geiger}
\affiliation{Physikalische Chemie, Universit\"at Dortmund,
  Otto-Hahn-Str. 6, D-44221 Dortmund, Germany}

\date{\today}

\begin{abstract}
We explore the phase diagram of TIP4P-Ew 
[J. Chem. Phys. {\bf 120}, 9665 (2004)] 
liquid model water from
the boiling-point down to $150\,\mbox{K}$ at densities
ranging from $0.950\,\mbox{g}\,\mbox{cm}^{-3}$ to
$1.355\,\mbox{g}\,\mbox{cm}^{-3}$.
In addition to the low-density/high-density (LDL/HDL) liquid-liquid transition,
we observe a high-density/very-high-density (HDL/VHDL) 
transformation for the lowest temperatures at $1.30\,\mbox{g}\,\mbox{cm}^{-3}$. 
A Van der-Waals type loop suggest the presence of a first order 
HDL/VHDL transition.
In addition, we identify a pre-transition 
at $1.24\,\mbox{g}\,\mbox{cm}^{-3}$, suggesting
the experimentally detected HDA/VHDA-transformation to be a two-step process.
For both pre- and main- HDA/VHDA-transition 
we observe a step-wise increase of the oxygen coordination
number for interstitial water molecules.
\end{abstract}

\maketitle

Experimental and theoretical studies
conducted over the past 14 years have provoked
the interpretation that water's anomalies 
stem from a transformation
between two major liquid forms of water buried in the
deeply supercooled region (see reviews by H.E. Stanley
and P.G. Debenedetti \cite{Debenedetti:2003:1,Debenedetti:2003}).
The two different liquids have their
counterparts in the glassy state: The
high density (HDA) and low density amorphous (LDA) ice forms
\cite{Mishima:84}.
However, it is still an open question how exactly the different
amorphous ice forms and supercooled liquid water
are connected, since the ``no man's land'' region
largely prohibits direct experimental access \cite{Debenedetti:2003}.
Therefore, starting with the work of P.H. Poole et al.\ in 1992,
basically  computer simulation studies have established a picture of
a first order liquid-liquid phase transition between two
liquids ending up in a metastable critical point
\cite{Poole:92,Tanaka:96:1}. Although singularity free scenarios
might as well explain the properties of 
supercooled water \cite{Debenedetti:2003},
there is experimental support for the liquid-liquid critical
point hypothesis from the changing slope of
the metastable melting curves
observed for different ice polymorphs \cite{Mishima:1998:2,Mishima:2000}.

Meanwhile, also
a very high density form (VHDA) of amorphous ice was observed and
shown to be distinct from HDA \cite{Loerting:2001}. 
Neutron scattering data reveals 
that the transformation between HDA and VHDA is related
to an increasing population of interstitial
water molecules in an O-O distance-interval between $3.1\,\mbox{\AA}$
and $3.3\,\mbox{\AA}$ \cite{Finney:2002}. 
Simulation studies indicate, that VHDA should be considered as 
the amorphous solid counterpart to the high density
liquid water phase at ambient conditions, 
and not HDA \cite{Giovambattista:2005:1,Giovambattista:2005:2}. 
Koza et al. \cite{Koza:2005} have demonstrated by using inelastic 
neutron scattering that HDA and VHDA appear
to be heterogeneous at the length-scale of nanometers
and that different forms of HDA are obtained, depending on
the exact preparation process \cite{Koza:2005}. Tulk et al.\ find
by annealing of HDA at normal pressure \cite{Tulk:2002} 
evidence for the presence
of a multitude of (metastable) amorphous ice states.
Finally, a very recent experimental study by Loerting et al. 
suggest the presence of a well established
first order phase transition between
the HDA and VHDA amorphous ices \cite{Loerting:2005}. 
In this context we would like to emphasize, that the
computer simulations of Brovchenko et
al. \cite{Brovchenko:2003,Brovchenko:2005} were the first
to conclude that there might exist even more than
one liquid-liquid transition in supercooled water. 

Here we present extensive thermophysical data
on the deeply supercooled liquid state of the 
TIP4P-Ew water model \cite{Horn:2004}.
TIP4P-Ew seems to be an ideal candidate, since it 
not only reproduces well the thermodynamic and structural properties
of the liquid phase \cite{Horn:2004} (including a boiling point at $370\,\mbox{K}$ \cite{Horn:2005}), but
the TIP4P family of models also describe 
amazingly well  the different ice polymorphs 
\cite{Sanz:2004,Vega:2005:1,Vega:2005:2}, qualitatively reproducing
the entire phase diagram of crystal solid water up to about $1\,\mbox{GPa}$. 
Recently, also attempts have been
undertaken to study the HDA to VHDA transition by computer simulations,
suggesting a continuous structural transformation \cite{Martonak:2005}.
However, facing the low temperature conditions, these studies, although
quite long on computer simulation time scales, suffer certainly from
equilibration problems.
To overcome these limitations, we present parallel tempering simulations
of an extended ensemble of states \cite{Marinari:1992}, applying
the technique of volume-temperature
replica exchange molecular dynamics simulation (VTREMD) 
\cite{Paschek:2005:1}. 
The Replica exchange molecular dynamics scheme (REMD), is a
parallel tempering variant \cite{Sugita:99}, based on
the MD simulation technique,
enhancing sampling by about 
two orders of magnitude \cite{Yamamoto:2000}.

For our simulation \cite{VTREMD} we consider a
grid of 644 ($V$,$T$)-states \cite{TIP4PSTATES}.
Starting from a set of equilibrated initial configurations obtained at
ambient conditions,
the imulation \cite{techdetails} was conducted for $50\,\mbox{ns}$,
providing a total $32.2\,\mu\mbox{s}$ worth of trajectory data.
The average time interval between two successful
state-exchanges was obtained to be about $3\,\mbox{ps}$. 
The overall shape of the phase diagram presented here is already
well established after only 
$2\,\mbox{ns}$ simulation time. However, since the configurational 
sampling of the lower-$T$ states happens almost only by exchange with
replicas coming from higher temperatures, the resolution of 
fine details of the $PVT$-diagram requires 
a long simulation run. The simulation took about 100 days 
on 20 Processors of our
Intel Xeon 3.05 GHz Linux cluster. All data reported here were averaged over
the final $45\,\mbox{ns}$ of the simulation.
\begin{figure}
  \centering
  \footnotesize
  \includegraphics[angle=0,width=6.5cm]{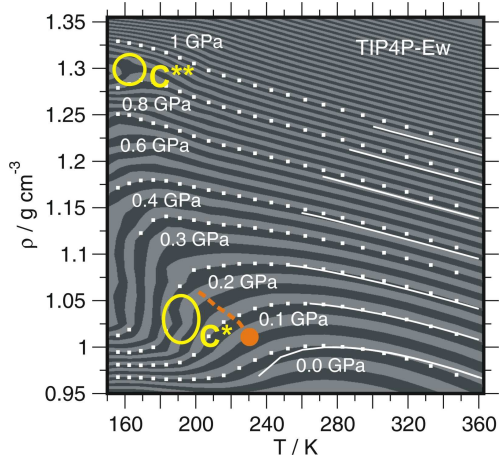}
  \caption{
    $P(V,T)$-surface of liquid TIP4P-Ew water \cite{PVTDATA}.
    The spacing between different contour colours corresponds to a pressure drop of 25 MPa.
    Selected isobars are indicated. The solid lines indicate the experimental
    isobars for $0.1\,\mbox{MPa}$, $0.1\,\mbox{GPa}$,
    $0.2\,\mbox{GPa}$, $0.4\,\mbox{GPa}$,
    $0.6\,\mbox{GPa}$, $0.8\,\mbox{GPa}$,
    and $1.0\,\mbox{GPa}$ \cite{Wagner:2002}.
    The ellipse indicates the
    as the LDL/HDL and HDL/VHDL critcal regions for TIP4P-Ew. 
    The large filled circle and the
    heavy dashed line indicates the metastable critical point and
    HDL/LDL-transition line for $\mbox{D}_2\mbox{O}$ according
    to Mishima \cite{Mishima:2000} projected on the TIP4P-EW 
    $P(\rho,T)$-surface.
    \label{fig:01}}
\end{figure}
\begin{figure}[!t]
  \centering
  \includegraphics[angle=0,width=5.6cm]{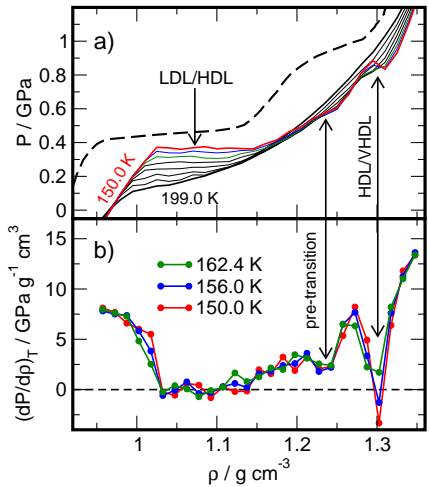}
  \caption{
    a) Isotherms $P(\rho,T)$ for the eight lowest temperatures.
       The dashed line represents the compression curve of amorphous ice
       at $125\,\mbox{K}$
       according to Loerting et al.\ \cite{Loerting:2005}.
    b) $\partial P(\rho,T)/\partial \rho)_T$ for 
    the three lowest temperature isotherms. The arrows indicate
    the location of the LDA/HDA and HDA/VHDA transitions, as well
    as the apparent pre-transition. 
    \label{fig:02}}
\end{figure}
\begin{figure}[!t]
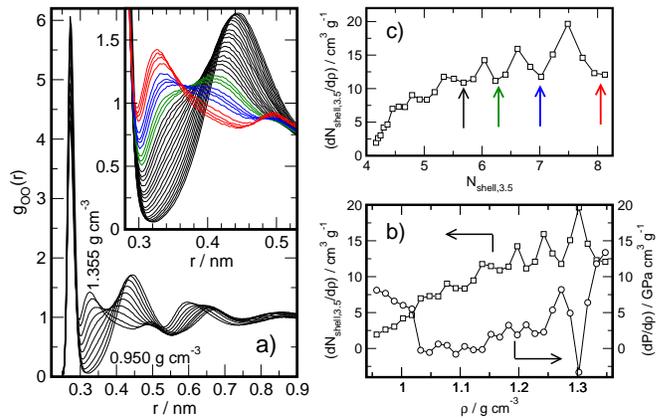

  \centering
  \includegraphics[angle=0,width=4.0cm]{fig03a.eps}
  \includegraphics[angle=0,width=4.5cm]{fig03bc.eps}
  \caption{
    a) OO radial pair distribution functions $g(r)$ for $150.0\,\mbox{K}$.
    Shown is every third density. The insert shows all OO-$g(r)$'s 
    for  the  ``interstitial region'' for $150.0\,\mbox{K}$. Changing colour 
    indicates gaps between the $g(r)$-lines.
    b) Derivative of the pressure with respect to the density for 
    the $150.0\,\mbox{K}$ isotherm and derivative of the water
    coordination number($r_{\rm OO}\!\leq\!0.35\,\mbox{nm}$). 
    respect to the density.
    c) Derivative of the water
    coordination number($r_{\rm OO}\!\leq\!0.35\,\mbox{nm}$). 
    respect to the density as a function of the coordination number.
    \label{fig:03}}
\end{figure}
\begin{figure*}[!t]
  \centering
  \includegraphics[angle=0,width=17.3cm]{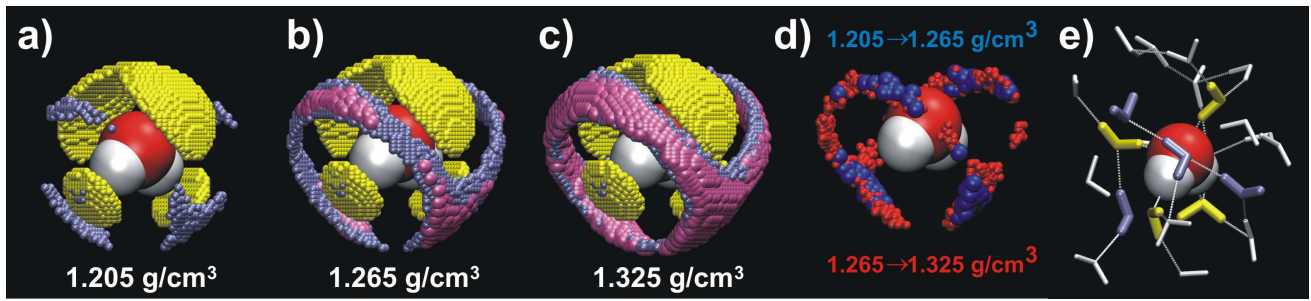}
  \caption{
    a,b,c)
    3D-distribution of water density around a central
    water molecule. Each small sphere represents a volume element
    with a water number density larger than $100\,\mbox{nm}^{-3}$ whereas the
    larger (mauve) spheres indicate densities exceeding then
    $130\,\mbox{nm}^{-3}$. Yellow
    indicates the first shell ($r\!\leq\!0.3\,\mbox{nm}$), whereas
    iceblue and mauve indicate intersticial 
    water ($r\!<\!0.3\leq 0.35\,\mbox{nm}$). Densities are indicated,
    $T\!=\!150\,\mbox{K}$.
    d) Volume elements showing a numer density increase of more
    than $45\,\mbox{nm}^{-3}$ when increasing the density
    $1.205\,\mbox{g}\,\mbox{cm}^{-3}\rightarrow1.265\,\mbox{g}\,\mbox{cm}^{-3}$
    (blue) and
    $1.265\,\mbox{g}\,\mbox{cm}^{-3}\rightarrow1.325\,\mbox{g}\,\mbox{cm}^{-3}$
    (red) at $150.0\mbox{K}$.
    e)
    Snapshot of a randomly selected solvation shell of a water molecule,
    $1.325\,\mbox{g}\,\mbox{cm}^{-3}$ at $150.0\,\mbox{K}$.
    Color coding is as in a.
    \label{fig:04}}
\end{figure*}

Figure \ref{fig:01} shows the phase diagram 
of liquid TIP4P-Ew water
as contour plot of the $P(\rho,T)$-data.
The TIP4P-Ew phase diagram exhibits a first
order low-density/high-density liquid-liquid phase 
transition, ending in a metastable critical point $C^{*}$. 
The ellipse denotes the region 
where LDL/HDL-critical point is apparently located. 
Due to a finer $T$-grid at lower temperatures, and longer
simulation times, we have improved the
sampling of the low-$T$ states, compared to our recent
study on the five-site TIP5P-E model \cite{Paschek:2005:1}. 
Figure \ref{fig:01} demonstrates that
TIP4P-Ew reproduces the experimental phase diagram of liquid water
much better than the TIP5P-E model \cite{Paschek:2005:1}, almost
quantitatively matching waters compressibility and thermal expansivity
up to pressures in the GPa-range.
Figure \ref{fig:01}a indicates, that the
Van der Waals-loop like overshooting of the $P(\rho,T)$-isotherm, as
observed in the HDL/LDL coexistence region of the
TIP5P-E model \cite{Paschek:2005:1},
is found to be almost completely flattened out in the case of TIP4P-Ew.
Also shown in Figure \ref{fig:01}
is the LDL/HDL-transition line and critical point
as obtained by O. Mishima \cite{Mishima:2000} for $\mbox{D}_2\mbox{O}$.
Since the densities of the experimental coexistence line are essentially
not known, we have projected the pressures reported by Mishima 
on the TIP4P-Ew $P(\rho,T)$-surface. We find a temperature difference 
between the two metastable critical 
points by about $30-40\,\mbox{K}$. We would like
to point out that this value corresponds roughly to the temperature difference
between the Ice-$I_h$ melting-lines 
of TIP4P-Ew and  $\mbox{D}_2\mbox{O}$ \cite{Vega:2005:1}.

In addition to the LDL/HDL-transition, we denote the appearance
of another (HDL/VHDL) 
liquid-liquid-transition at  $1.30\,\mbox{g}\,\mbox{cm}^{-3}$
at about $165\,\mbox{K}$. The second metastable critical region,
denoted by $C^{**}$, is identified by a van der Waals
type of behavior of $P$ vs. $\rho$ shown in Figure \ref{fig:02}a. 
Moreover, Figure \ref{fig:02}b presents the derivative $(\partial P(\rho,T)/\partial \rho)_T$
of the lower-$T$ isotherms characterizing the HDL/VHDL-transition by
a pronounced dip. In addition to the fully developed first order transition
at $1.30\,\mbox{g}\,\mbox{cm}^{-3}$, we would like to point out
the appearence of a shoulder at about  $1.24\,\mbox{g}\,\mbox{cm}^{-3}$
(Figure \ref{fig:02}a), corresponding to a less well
pronounced dip in Figure \ref{fig:02}b.
Both curves indicate the emergence of a not yet fully developed transition
around $1.24\,\mbox{g}\,\mbox{cm}^{-3}$,
which we refer to as the HDL/VHDL {\em pre-transition}.
We find that our simulations are in qualitative agreement
with the experimental compression curve of amorphous water obtained 
at $125\,\mbox{K}$ shown in Figure \ref{fig:02}a. However,
there is a shift of about $0.06\,\mbox{g}\,\mbox{cm}^{-3}$ between
the experimental and simulated data-sets. 
The elastic compression of VHDA as obtained by Loerting et al.\ with  
$0.10\,\mbox{g}\,\mbox{cm}^{-3}\,\mbox{GPa}^{-1}$ \cite{Loerting:2005}
is comparable to the $0.08\,\mbox{g}\,\mbox{cm}^{-3}\,\mbox{GPa}^{-1}$
(Figure \ref{fig:02}b) of the VDHL-phase observed here.
The {\em larger} elastic compression found for
HDA ($\approx 0.14\,\mbox{g}\,\mbox{cm}^{-3}\,\mbox{GPa}^{-1}$
\cite{Loerting:2005}) is also qualitatively in accordance with 
our simulation data.
Moreover, the slightly 
asymmetric shape of HDA/VHDA the transition observed by Loerting 
et al.\ \cite{Loerting:2005} over the density interval 
between $1.18\,\mbox{g}\,\mbox{cm}^{-3}$
$1.30\,\mbox{g}\,\mbox{cm}^{-3}$ does not seem to 
rule out the presence of a softer transition or pre-transition in the
density range of $1.22\,\mbox{g}\,\mbox{cm}^{-3}$ to
$1.24\,\mbox{g}\,\mbox{cm}^{-3}$ and a sharper transition at
about $1.28\,\mbox{g}\,\mbox{cm}^{-3}$.

Figure \ref{fig:03} shows how the observed
transitions are related to structural changes
in the local environment of the water molecules. 
Figure \ref{fig:03}a depicts the O-O pair 
correlation functions for water at
$150\,\mbox{K}$ as a function of density. The insert given
in Figure \ref{fig:03}a focuses primarily
on the changes in the so
called interstitial water region between 
$0.3\,\mbox{nm}$ and $0.35\,\mbox{nm}$. In line with
structural data obtained from neutron 
scattering experiments \cite{Finney:2002},
we observe the appearence of a peak in the 
interstitial region as pressure increases.
Moreover, with increasing density we denote the occurrence
{\em undulations} in the water coordination number.
In the insert in Figure \ref{fig:03}a these undulations appear as 
larger gaps between the individual $g(r)$-curves. 
For better visibility these increased gaps are indicated by
a changing color. The magnitude of the variation in the local density is
shown in Figure \ref{fig:03}b by plotting
derivative of waters O-O coordination
number for $r\leq 0.35\,\mbox{nm}$  with respect to the density
$(\partial N_{3.5}/\partial\rho)_T$. Comparing this data with
the $(\partial P(\rho,T)/\partial \rho)_T$-line 
(Figure \ref{fig:03}b) we find that the observed
HDL/VHDL transitions are exactly accompanied
by a step-wise increase of water's coordination
number. 
Moreover,  Figure \ref{fig:03}b presents the
 $(\partial N_{3.5}/\partial\rho)_T$ data
as a function of the coordination number $N_{3.5}$. The diagram
indicates that the HDL/VHDL transition is accompanied by an increase
from seven to eight water molecules in the sphere of 
$0.35\,\mbox{nm}$ around a central water molecule.
At the HDL/VHDL pre-transition we find only 
a fractional increase of about $0.75$
water molecules on average. Further wiggles in 
$(\partial N_{3.5}/\partial\rho)_T$ suggest
that there is at least one more step in the occupation
number at even lower densities.

Figure \ref{fig:04} illustrates the structural changes 
of the local water environment along 
the HDL/VHDL pre- and main-transitions. In line with
experimental finding, waters first coordination shell 
($r\leq0.3\,\mbox{nm}$) is found to consist of four
tetrahedrally arranged water molecules \cite{Finney:2002}. 
With increasing pressure, water is penetrating 
the interstitial region
($0.3\,\mbox{nm}\leq r\leq0.35\,\mbox{nm}$) arranging 
in a lobes around the positions occupied by the first-shell
water molecules. However, even at the highest densities there
are not more than about four water molecules filling the interstitial
positions, indicating that the lobe-type positions can 
be only partially occupied. Figure \ref{fig:04}a indicates
that for HDL-water the interstitial molecules are 
preferentially located on the ``anti-tetrahedral'' positions,
being located on the centers of faces of the tetrahedron formed
by the first-shell water molecules.
Figure \ref{fig:04}b and \ref{fig:04}d 
indicate that along the pre-transition
interstitial water is also predominantly inserting into
these ``anti-tetrahedral'' positions, whereas during
the HDL/VHDL main-transition the lobes are getting more and
more completed by increasingly filling the gaps
between two ``anti-tetrahedral'' sites.
Figure \ref{fig:04}e exemplifies this scenario, showing
four nearest neighbor water molecules and three water 
molecules in ``anti-tetrahedral''-coordination. The water
molecule representing on the connecting lobe-position 
is  apparently forming a hydrogen bond-bridge between the 
water molecules. 

In conclusion, our simulations indicate the presence of a
major HDL/LDL transition in supercooled water, being
responsible for the anomalous thermodynamical behavior of
water at ambient conditions. However with decreasing
temperature, water's $P(\rho,T)$ surface becomes 
increasingly corrugated, developing further
step-like transformations: The HDL/VHDL pre- and main-transitions,
and possibly more at even lower temperatures.
The onset of these transformations is
apparently tightly related to a change of the local coordination 
of the water molecules.
Hence we propose a two-region scenario for 
supercooled-water: A very low-temperature
region with several distinct structural transitions, which is
typically probed by studies on the amorphous forms
of water \cite{Loerting:2001,Loerting:2005,Koza:2005,Tulk:2002},
and a medium temperature region, where
due to increased thermal energy these transitions are covered
and two major liquid forms of water (LDL and HDL) are 
found to be in equilibrium.

We gratefully acknowledge support by the 
Deutsche Forschungsgemeinschaft (FOR 436).


\begin{thebibliography}{32}
\bibitem[{VTR()}]{VTREMD}
\bibinfo{note}{State swapping moves between two states $i$ and $j$ are accepted
  with a probability $P_{acc} \mbox{min} \{ 1,\exp[ \beta_i \left(
  U(\vec{s}_i^N;L_i) - U(\vec{s}_j^N;L_i) \right) + \beta_j \left(
  U(\vec{s}_j^N;L_j) - U(\vec{s}_i^N;L_j) \right) ]\}$. Here, $\vec{s}_i^N$
  represents the set of scaled coordinates $\vec{s}_N\!=\!L^{-1}\,\vec{r}_N$ of
  the entire $N$-particle system belonging to state $i$. $U(\vec{s}_i^{N};L_i)$
  denotes the potential energy of configuration $\vec{s}_i^N$ at volume
  $V_i=L_i^3$, whereas $U(\vec{s}_i^{N};L_j)$ represents the configurational
  energy belonging to $\vec{s}_i^N$ at volume $V_j$. Whether a state swapping
  move or an MD move is executed, is chosen at random with a probability of
  $0.2$ for selecting state swapping moves.}
\end{thebibliography}
\end{document}